# Standard Model Baryogenesis

M.B. Gavela, P.Hernández, J. Orloff and O.Pène

Contribution to the XXIXth Rencontre de Moriond,
"Electroweak Interactions and Unified Theories".

Presented by M. B. Gavela
CERN, TH Division, CH-1211, Geneva 23, Switzerland

**Abstract**

Simply on CP arguments, we argue against a Standard Model explanation of baryogenesis via the charge transport mechanism. A CP-asymmetry is found in the reflection coefficients of quarks hitting the electroweak phase boundary created during a first order phase transition. The problem is analyzed both in an academic zero temperature case and in the realistic finite temperature one. At finite temperature, a crucial role is played by the damping rate of quasi-quarks in a hot plasma, which induces loss of spatial coherence and suppresses reflection on the boundary even at tree-level. The resulting baryon asymmetry is many orders of magnitude below what observation requires. We comment as well on related works.

# 1 Introduction

The baryon number to entropy ratio in the observed part of the universe is estimated to be $n_B/s \sim (4-6)10^{-11}$[1] . In 1967, A.D. Sakharov [2] established the three building blocks required from any candidate theory of baryogenesis: a) Baryon number violation, b) C and CP violation, c) Departure from thermal equilibrium.

The Standard Model (SM) contains a)[3] and b)[4], while c) could also be large enough [5][6], if a first order $SU(2) \times U(1)$ phase transition took place in the evolution of the universe [7]. An explanation within the SM would be a very economical solution to the baryon asymmetry puzzle. Unfortunately, intuitive arguments lead to an asymmetry many orders of magnitude below observation [8][9]. However, the study of quantum effects in the presence of a first order phase transition is rather delicate, and traditional intuition may fail. The authors of ref.[10] have recently studied this issue in more detail and claim that, in the finite temperature charge transport mechanism [11], the SM is close to produce enough CP violation as to explain the observed $n_B/s$ ratio. In this talk, we summarize our study [12] of the Standard Model C and CP effects in an electroweak baryogenesis scenario. Even if one assumes an optimal sphaleron rate and a strong enough first order phase transition, we discard this scenario as an explanation of the observed baryon number to entropy ratio.

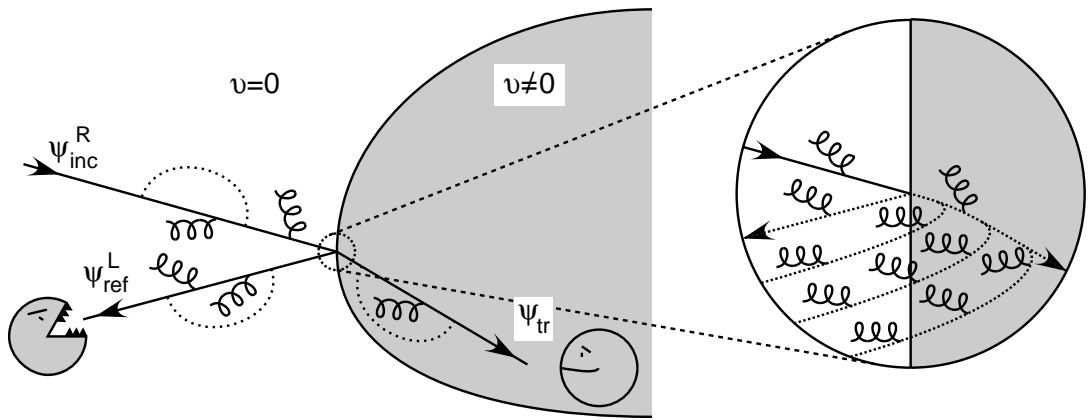

Figure 1: *Artistic view of the charge transport mechanism, as described in the text. The hungry "pacman" represents rapid sphalerons processes. The wiggly lines stand for collisions with thermal gluons in the finite temperature case; they are absent in the academic $T = 0$ model. Only electroweak loops are depicted, represented by dotted lines.*

A first order phase transition can be described in terms of bubbles of "true" vaccuum (with an inner vaccuum expectation value of the Higgs field $v \neq 0$) appearing and expanding in the preexisting "false" vaccuum (with $v = 0$ throughout). We can "zoom" into the vicinity of one of the bubbles. There the curvature of its wall can be neglected and the world is divided in two zones: on the left hand side, say, $v = 0$; on the right $v \neq 0$. The actual bubble expands from the broken phase ($v \neq 0$) towards the unbroken one ($v = 0$). We work in the wall rest frame in which the plasma flows in the opposite direction. Consider thus a baryonic flux hitting the wall from the unbroken phase. The heart of the problem lies in the reflection and transmission properties of quarks bumpimg on the bubble wall. CP violation distinguishes particles from antiparticles and it is *a priori* possible to obtain a CP asymmetry on the reflected baryonic current, $\Delta_{CP}$. The induced baryon asymmetry is at most $n_B/s \sim 10^{-2}\Delta_{CP}$, in a very optimistic estimation of the non-CP ingredients [13][10].

The symmetries of the problem have been analyzed in detail [14] for a generic bubble. The analytical results correspond to the thin wall scenario. The latter provides an adequate physical description for typical momentum of the incoming particles $|\vec{p}|$ smaller than the inverse wall thickness $l$, i.e., $|\vec{p}| \ll 1/l$. For higher momenta, cutoff effects would show up, but it is reasonable to believe that the thin wall approximation produces an upper bound for the CP asymmetry. We work in a simplified scenario with just one spatial direction, perpendicular to the wall surface: phase space effects in the $3+1$ dimension case would further suppress the effect.

The precise questions to answer in the above framework are : 1) the nature of the physical process in terms of particles or quasi-particles responsible for CP violation, 2) the order in the electroweak coupling constant, $\alpha_W$, at which an effect first appears, 3) the dependence on the quark masses and the nature of the GIM cancellations involved.

We shall consider the problem in two steps: zero temperature scenario ($T = 0$) in the presence of a wall with the non-equilibrium situation mimicked by assuming a flux of quarks hitting the boundary from just one phase, and finite temperature case. Intuition indicates that an existing CP violating effect already present at zero temperature will diminish when the system is heated because the effective v.e.v. of the Higgs field decreases and in consequence the fermion masses do as well (only the Yukawa couplings already present at $T = 0$ remain unchanged). This intuition can be misleading only if a new physical effect, absent at $T = 0$ and relevant for the problem, appears at finite temperature. We discuss and compare the building blocks of the analysis in both cases. The $T = 0$ case provides a clean analysis of the novel aspects of the physics in a world with a two-phase vacuum.

At finite temperature, a plasma is an incoherent mixture of states. CP violation is a quantum phenomenon, and can only be observed when quantum coherence is preserved over time scales larger than or equal to the electroweak time scales needed for CP violation. This is however not the case in the plasma, where the scattering of quasi-quarks with thermal gluons induces a large damping rate, $\gamma$.

We show that tree-level reflection is suppressed for any light flavour by a factor $\sim m/2\gamma$. The presently discussed CP-violation observable results from the convolution of this reflection effect with electroweak loops in which the three generations must interfere coherently in order to produce an observable CP-violation. It follows that further factors of this type appear in the final result, which is many orders of magnitude below what observation requires and has an "à la Jarlskog"[8] type of GIM cancellations.

The results of our analysis indicate that in the presence of a first order phase transition, a CP-asymmetry in the SM appears at order $\alpha_W^2$, has a conventional type of GIM cancellation and chiral limit, and it is well below what observation requires in order to solve the baryon asymmetry puzzle.

## 2  Zero temperature

The necessary CP-odd couplings of the Cabibbo-Kobayashi-Maskawa (CKM) matrix are at work. Kinematic CP-even phases are also present, equal for particles and antiparticles, which interfere with the pure CP-odd couplings to make them observable. These are the reflection coefficients of a given particle hitting the wall from the unbroken phase. They are complex when the particle energy is smaller than its (broken phase) mass. Finally, as shown in [12] [14] , the one loop self-energy of a particle in the presence of the wall cannot be completely renormalized away and results in physical transitions. Such an effect is absent for on-shell particles in a world with just one phase. The difference is easy to understand: the wall acts as an external source of momentum in the one-loop process. The transitions between any two flavors of the same charge produce a CP violating baryonic flow for any given initial chirality.

The essential non-perturbative effect is the wall itself. The propagation of any particle of the SM spectrum should be exactly solved in its presence. And this we do for a free fermion, leading to a new Feynman propagator which replaces and generalizes the usual one. The propagator for quarks in the presence of the wall contains massless and massive poles:

$$S(q^f, q^i) = -1/2 \Bigl\{ \frac{1}{q_z^f - q_z^i + i\epsilon} \left( \frac{1}{\slashed{q}^f} + \frac{1}{\slashed{q}^i} \right) - \frac{1}{q_z^f - q_z^i - i\epsilon} \left( \frac{1}{\slashed{q}^f - m} + \frac{1}{\slashed{q}^i - m} \right) +$$
$$\frac{1}{\slashed{q}^f - m} \gamma_z \frac{1}{\slashed{q}^i} - \frac{1}{\slashed{q}^f} \gamma_z \frac{1}{\slashed{q}^i - m} -$$
$$\frac{m}{\slashed{q}^f(\slashed{q}^f - m)} \left[ 1 - \frac{m \gamma_0}{E + p'_z}(1 - \alpha_z) \right] \gamma_0 \frac{m}{\slashed{q}^i(\slashed{q}^i - m)} \Bigr\} \quad (1)$$

where we have assumed for simplicity zero momentum parallel to the wall ($q_x^i = q_y^i = q_x^f = q_y^f = 0$). Due to the wall the initial and final $z$ components of the momentum need not be equal. All denominators in the usual Feynman propagators in eq. (1) should be understood as containing a supplementary $+i\epsilon$ factor. Besides this traditional source of phases, the propagator contains new CP-even ones in $p'_z = \sqrt{E^2 - m^2}$, which becomes imaginary in the case of total reflection ($E < m$ where $E$ is the fermion energy).

With this exact, non-perturbative tool, perturbation theory is then appropiate in the gauge and Yukawa couplings of fermions to bosons, and the one loop computations can be performed. Strictly speaking the gauge boson and Higgs propagators in the presence of the wall are needed, and it is possible to compute them with a similar procedure [15]. In particular this implies to consider loops with unbroken, broken and mixed contributions. For the time being, we work in a simplified case in which the wall does not act inside quantum loops. These are computed in the broken phase. We considered one-loop electroweak effects which bring the CKM phase into the game. A toy computation indicates that a negligible CP-asymmetry first appears at order $\alpha_W$ in amplitude, with two unitarity triangles describing the type of GIM cancellations of the problem [14]. For a thin wall the non-local character of the internal loop is important because large particle momenta $\sim M_W$ are present and $l \ll M_W^{-1}$. Our calculation suggests that an even smaller result (although always at the same electroweak order) would follow for a more realistic thick wall, $l \gg M_W^{-1}$, where a local appoximation could be pertinent.

## 3   Non zero temperature

The three building blocks are analogous to the $T = 0$ case: CKM CP violation, CP-even phases in the reflection coefficients [10] and the fact that the fermion self-energy at finite $T$ results in physical transitions.

A fundamental difference with the $T = 0$ case is the damping rate, $\gamma$, of quasi-particles in a plasma. Due to incoherent thermal scattering with the medium, their energy and momentum are not sharply defined, but spread like a resonance of width $2\gamma$ [16]. The quasi-particle has thus a finite life-time, turning eventually into a new state, out of phase with the initial one. Small momenta are relevant for the problem under study, and it is known that at zero momentum the QCD damping rate is of the order $\gamma \sim 0.15 g_s^2 T$ [16], i.e. $\sim 19$ GeV at $T = 100$ GeV. Although the imaginary part of the QCD self-energy is smaller than its real part [17]–[20], which settles the overall scale of the quasi-particle "masses", it is much larger than the real part of the electroweak self-energy. It should weaken the effect of electroweak level splitting, essential to the asymmetry.

A first step is the computation of the spectrum. The on-shell states correspond to the zeros of the determinant of $i\gamma_0(S^{-1})$, and the corresponding eigenstates verify the effective Dirac equation

$$\begin{pmatrix} -i\partial_t - \tfrac{1}{3} i\sigma_z \partial_z - i\gamma + \omega_R^0 & \tfrac{m}{2}\theta(z) \\ \tfrac{m}{2}\theta(z) & -i\partial_t + \tfrac{1}{3} i\sigma_z \partial_z - i\gamma + \omega_L^0 \end{pmatrix} \psi(z,t) = 0 \qquad (2)$$

where $\omega_R^0$, $\omega_L^0$ are the zero momentum energies of the left/right quasiparticles in the unbroken phase. Notice that the group velocity of the solutions of eq. (2) has been approximated by $1/3$. The spectrum far from the wall in both the unbroken and broken phase is sketched in fig. 2.

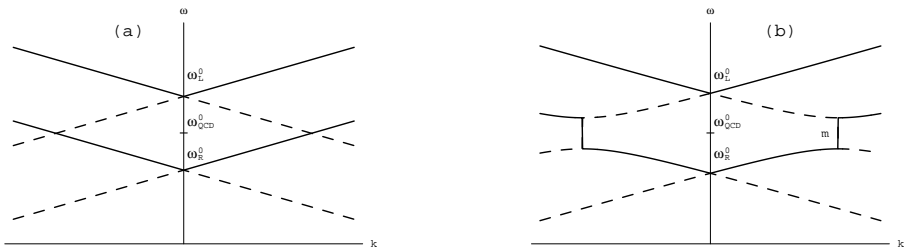

Figure 2: *Dispersion relations for quasi-particles in the (a) unbroken and (b) broken phases. The full (dashed) lines are normal (abnormal) branches. The upper (lower) lines correspond to left (right) chirality. The vertical lines in (b) represent the gaps of width $\simeq m$, in which total reflection occurs.*

Let us start by considering the one flavor case. The value of the damping rate, $\gamma$, and the uncertainty principle imply to describe the incoming quasiparticles as wave packets whose size $d$ cannot exceed the mean free path $\sim 1/6\gamma$. The effective

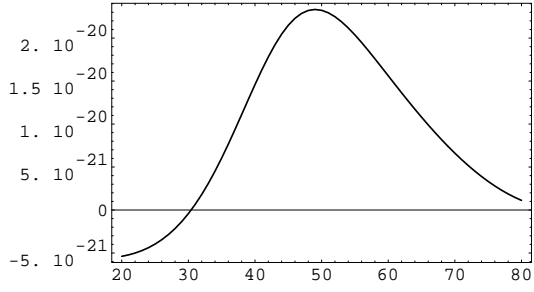

Figure 3: The dominant CP asymmetry when mass effects are included inside thermal loops, as a function of the energy. It corresponds to charge 2/3 flavors and appears at order $(O(\alpha_W^2))$.

Dirac equation (2) determines then the time evolution of the wave packet. Since eq. (2) has a non hermitean part proportional to $\gamma$, the total probability of our wave packets falls off exponentially in time. This reflects the fact that quantum coherence is lost after a time $\sim 1/2\gamma$. This loss of probability is compensated by a continuous probability of creation of new wave packets, tuned so as to keep constant the total particle density and thus preserve unitarity. The new wave packets are assumed to be produced with a random phase, i.e. out of phase of the wave packets that have decayed. This expresses the fact that thermal collisions with the plasma incoherently turn the quasiparticle into a new one with a different energy and momentum spectrum.

The details of the derivation are described in [22]. We estimate the reflection probability of the incoming wave packet close to the wall, i.e. the flux of particles reflected back by the wall into the unbroken phase per unit incoming flux. For a chirality $\chi$ ($L$ or $R$) we define a reflection coefficient, function of the energy:

$$r_\chi(\omega) = -\frac{m/2}{p(\omega) + e^{i\phi}\sqrt{|p(\omega)^2 - m^2/4|}}, \qquad (3)$$

with

$$\phi = \frac{\arg(p(\omega) - m/2) + \arg(p(\omega) + m/2)}{2}, \quad p(\omega) = \omega - \frac{\omega_\chi^0 + \omega_{-\chi}^0}{2}, \qquad (4)$$

Notice that the reflection coefficient (3) becomes complex for an energy range of width $m$: $-m/2 < p(\omega) < m/2$, corresponding to total reflection.

Using Gaussian wave packets and the analyticity of the functions $r_\chi(\omega)$ and via a formal extension of the $t_0$-integral to $+\infty$, it is possible to show [22] that

$$n_r(0,0) = \int dE_0\, n_F(E_0) \left[ \int d\omega \frac{3d}{\sqrt{\pi}} e^{-9d^2(\omega - E_0)^2} |r(\omega + i\gamma)|^2 - \alpha m^2 (3d)^3 2\gamma \right] \qquad (5)$$

for $m \ll \gamma \ll 1/3d$. $\alpha$ varies from 0 to $(\pi - 2)/8\pi^{-1/2}$ depending on the importance of the would-be $t_0 > 0$ contribution. The last term in eq. (5) can be neglected.

In this limit the reflected density is thus a gaussian smear-out of $|r(\omega + i\gamma)|^2$, with a maximum value $|r_{max}|^2 = m^2/16\gamma^2$, instead of 1 when $\gamma = 0$. One

way to understand the physical origin of this reduction is to notice that, while the quasi-particles in the plasma are widely spread in energy and momentum, $d^{-1} \gg 6\gamma \gg m$, reflection (i.e. CP-even phases) is only important in a very narrow energy band, $\delta\omega \sim m$. Hence quasi-particles can hardly be reflected, but for the top flavor. In other words, it takes the wall a long time ($\sim 1/m$) to emit the reflected component of a small incoming packet. If the packet decays rapidly in a time $\sim 1/2\gamma$, it is natural to see the reflected wave strongly depleted by a factor $\sim m/2\gamma$.

Now we turn to several flavors and compute the CP asymmetry. Using the following values for the masses in GeV, $M_W = 50$, $M_Z = 57$, $m_d = 0.006$, $m_s = 0.09$, $m_b = 3.1$, $m_u = 0.003$, $m_c = 1.0$, $m_t = 93.7$, the couplings [12] $\lambda_d = 1.2\,10^{-4}$, $\lambda_s = 1.8\,10^{-3}$, $\lambda_b = 6.2\,10^{-2}$, $\lambda_u = 6.2\,10^{-5}$, $\lambda_c = 2\,10^{-2}$ and $\lambda_t = 1.88$, and $\alpha_s = 0.1$, $\alpha_W = 0.035$ we obtain for the integrated asymmetry,

$$\frac{\Delta_{CP}^{uct}}{T} = 1.6\,10^{-21}, \qquad \frac{\Delta_{CP}^{dbs}}{T} = -3\,10^{-24}. \tag{6}$$

In both cases the asymmetry is dominated by the two heavier external quarks. The induced baryon asymmetry $n_b/s$ cannot exceed $10^{-2}$ times [10] these results.

Fig. 3 shows $\Delta(\omega)$ for up quarks.

In ref. [10] Farrar and Shaposhnikov (FS) obtain $\Delta_{CP}/T \gtrsim 10^{-8}$, and conclude $n_B/s \sim 10^{-11}$ (see eq. (10.3) in [10]). Their result is many orders of magnitude above ours, eq. (6). The main origin of the discrepancy is that they have not considered the effect of the damping rate on the quasi-particle spectrum[1].

For the sake of comparison, we consider their approximation, i.e., with just the unbroken phase inside the thermal loops, both with zero and non zero damping rate, for a thin wall. In the energy region where the maximum asymmetry was found for $\gamma = 0$ [10] and down quarks, the $\alpha_W$ expansion with non zero damping rate leads to:

$$\Delta(\omega) = \left[\sqrt{\frac{3\pi}{2}}\frac{\alpha_W T}{32\sqrt{\alpha_s}}\right]^3 J \frac{(m_t^2 - m_c^2)(m_t^2 - m_u^2)(m_c^2 - m_u^2)}{M_W^6} \frac{(m_b^2 - m_s^2)(m_s^2 - m_d^2)(m_b^2 - m_d^2)}{(2\gamma)^9} \tag{7}$$

where $J = c_1 c_2 c_3 s_1^2 s_2 s_3 s_\delta$. This result shows the expected GIM cancellation and regular chiral behaviour. Its magnitude, $\sim 4\,10^{-22}$, is lower than the dominant one at order $\alpha_W^2$, shown in fig. 4(b).

Furthermore, we confirm the validity of their numerical calculation with zero damping rate, with no $\alpha_W$ expansion involved and as can be seen in fig. 4(a). The same computation including the damping rate is also shown in fig. 4(b).

A final comment on the wall thickness $l$ is pertinent. The mean free path for quasi-particles of lifetime $\sim 1/2\gamma$ and group velocity $1/3$ is $1/6\gamma \sim (120\,GeV)^{-1}$. The thin wall approximation is valid only for $l \ll 1/6\gamma$, while perturbative estimates[10] give $l \gtrsim (10\,GeV)^{-1} \gg 1/6\gamma$. A realistic $CP$ asymmetry generated in such scenario will be orders of magnitude below the thin wall estimate in eq. (6), reinforcing thus our conclusions, because a quasi-particle would then collide and loose coherence long before feeling a wall effect. This caveat should also be considered in any non-standard scenario of electroweak baryogenesis, where the wall thickness is larger than the mean free path.

---

[1]More precisely, they take into account the finite mean free path of the quasi-particles in the suppression factor, i.e. what fraction of the $\Delta_{CP}$ is transformed into a baryon asymmetry by the sphalerons, but not in the computation of $\Delta_{CP}$.

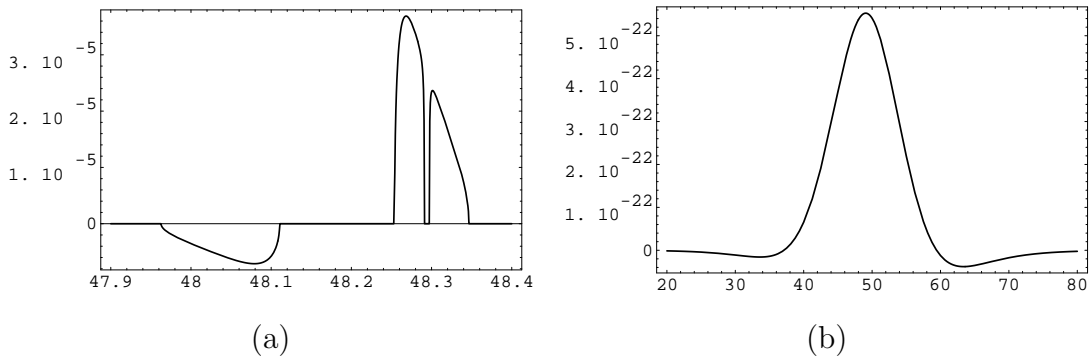

Figure 4: (a) shows the CP asymmetry produced by down quarks in the narrow energy range which dominates for zero damping rate, when masses are neglected in the internal loop. (b) shows the dramatic effect of turning on the damping rate effects, in the same approximation.

We have enjoyed several enlightening discussions with Tanguy Altherr and we express our sorrow for his sudden death. We acknowledge Luis Alvarez-Gaumé, Philippe Boucaud, Gustavo Branco, Andy Cohen, Alvaro De Rújula, Savas Dimopoulos, Jean Marie Frère, Jean Ginibre, Gian Giudice, Patrick Huet, Jean-Pierre Leroy, Manolo Lozano, Jean-Yves Ollitreault, Carlos Quimbay, Anton Rebhan, Eric Sather and Dominique Schiff for many fruitful discussions. Pilar Hernandez acknowledges partial financial support from NSF-PHY92-18167 and the Milton Fund. This work was supported in part by the Human Capital and Mobility Programme, contract CHRX-CT93-0132.

## 4  Note added

After the Moriond Conference, Huet and Sather [21] have analyzed the finite temperature problem. These authors state that they confirm our conclusions. As we had done in ref. [12], they stress that the damping rate is a source for quantum decoherence, and use as well an effective Dirac equation which takes it into account. They discuss a nice physical analogy with the microscopic theory of reflection of light. They do not use wave packets to solve the scattering problem, but spatially damped waves. Subsequently, we have submitted for publication two lengthy papers containing the details of our computations [14] [22].

In a recent note [23] Farrar and Shaposhnikov (FS) have expressed doubts on the technical reliability of both our work and that of Huet and Sather [21]. They claim that our schemes violate unitarity. This is incorrect, as particle number is always conserved in our approaches. The effective Dirac equation for a given quasi-particle contains indeed an imaginary component which parametrizes the damping rate. An effective description of the evolution of a subsystem of a larger entity does not have to be hermitian. In fact, consistency may imply an apparent lack of unitarity in a subensemble of a whole unitary system. We had explicitly discussed this point in the detailed version of our results [22]. We developed there a density matrix formalism containing a creation term of quasi-particles due to collisions with the medium, which exactly compensates disappearance by the same processes, see eq. (4.30) in [22]. The density of quasi-particles is always normalized to the equilibrium density. The quasi-particles created by the medium are out of phase with respect to the ones destroyed, but total particle number, and thus, unitarity, is preserved. FS also object to our claim that the reflection amplitude is suppressed when integrated over a wave packet by the interference between the contribution of different momenta. They state that, as the reflection phase shift varies between 0 and $\pi$, the different contributions should still sum up to a significant result. This argument fails as the total phase to consider is the combination of the above mentioned one with the optical path length, i.e. the phase of the $e^{ipz_0}$ factors, as may be seen in section 4.1.2 in [22]. Consequently, the phase shifts range between 0 and $2\pi$ resulting in a very strong destructive interference. The rest of the note by FS contains either comments which are unrelated to the main point under discussion, or unproved speculations on alternative scenarios. In fact, these authors have not demonstrated their implicit claim that the damping rate is irrelevant to the problem, neither have they proven us or Huet and Sather wrong in any concrete point of the calculations. And they have not included the effects of the damping rate in their explicit computation of the reflection properties. There is no point in being repetitive, and we refer the interested reader to the published work [12] [21] [14] [22].

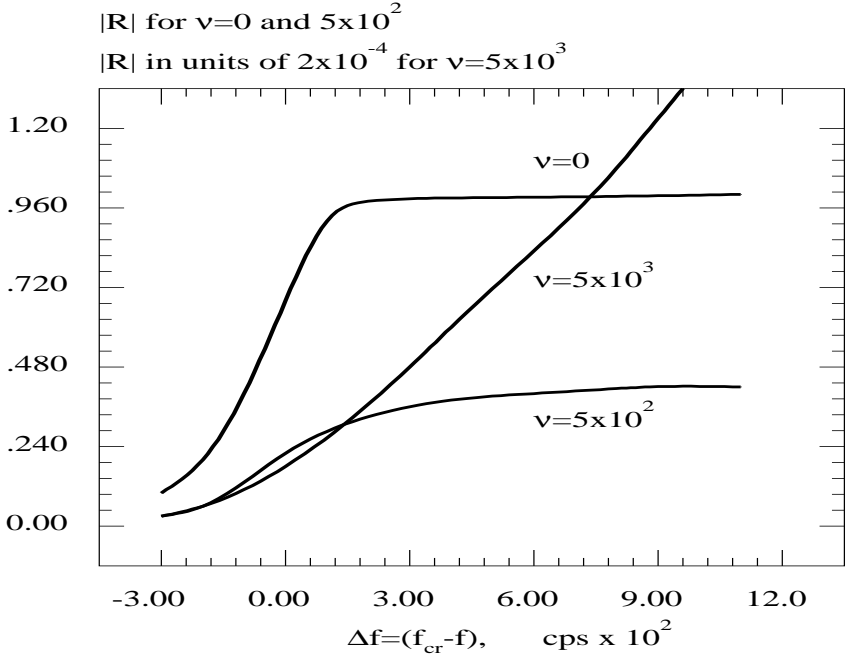

Figure 5: *Coefficient of reflection $|R|$ for the parabolic layer $z_m = 120km$, $\lambda_k = \frac{c}{f_k} = 30m$ for various values of the number of collisions $\nu$.*

It may be natural to wonder wether an analogous suppression of the reflection probability by damping phenomena has been encountered elsewhere in plasma physics. We did not find in the litterature a completely analogous scenario, and any new problem deserves a new analysis. However, some general trends have been found, for instance in the case of electromagnetic waves propagating in non-homogeneous plasmas. We recall such an example as an illustrative guideline to this type of physics, rather than as an argument in the present discussion. In ref. [24], several examples of the reflection of electromagnetic waves on the boundaries separating regions of different dielectric properties are analyzed, both with and without absorption. Strong reflection occurs whenever the wave has to cross a region of negative dielectric constant, where it is spatially damped. With a proper translation, it is easy to see that our wave equation in the presence of a thin wall is equivalent to the equation of the electromagnetic wave in the presence of the "transitional" layer in [24]. There, the effect of absorption is parametrized through an imaginary part in the dielectric constant. They explicitly compare the reflection coefficient for a layer of parabolic shape, with and without absorption, see Fig.5 taken from [24]. We present this figure although the physical situation differs sensibly from ours in that their total photon density is not constant. Nevertheless it illustrates the damping of the reflection coefficient due to incoherent interactions.

The horizontal axis measures the difference between the frequency of the electromagnetic wave and the critical frequency, for which the dielectric constant becomes negative and reflection is strong. $\nu$ is the collision frequency responsible for absorption. Notice the dramatic damping of the reflection coefficient for non-zero $\nu$: the curve corresponding to $\nu = 5 \cdot 10^3$ is scaled up by a factor $2 \cdot 10^4$.